\documentclass{article}

\usepackage{amsfonts}
\usepackage{amsmath}
\usepackage{amssymb}
\usepackage{color}
\usepackage{bm}
\usepackage{natbib}

\bibpunct{(}{)}{;}{a}{,}{,}

\begin{document}

\title{A Comparison of Methods of Inference in Randomized Experiments from a Restricted Set
of Allocations}
\author{Junni L. Zhang\footnote{Corresponding Author. Postal address: National School of Development, Peking University, 5 Yiheyuan Road, Haidian District, Beijing, 100871, China. Email: junnizhang@163.com.}\\
National School of Development, \\
Center for Statistical Science and Center for Data Science\\
Peking University, China\\
\\
and
Per Johansson\\
Department of Statistics, Uppsala University, Sweden}
\date{}
\maketitle

\begin{abstract}
Rerandomization is a strategy of increasing efficiency as compared to complete randomization. The idea with rerandomization is that of removing allocations with imbalance in the observed covariates and then randomizing within the set of allocations with balance in these covariates. Standard asymptotic inference based on mean difference estimator is however conservative after rerandomization. Given a Mahalanobis distance criterion for removing imbalanced allocations, \citet{Li&etal:2018} derived the asymptotic distribution of the mean difference estimator and suggested a consistent estimator of its variance. This paper discusses several alternative methods of inference under rerandomization, and compare their performance with that of the method in \citet{Li&etal:2018} through a large Monte Carlo simulation.  We conclude that some of the methods work better for small or moderate sample sized experiments than the method in \citet{Li&etal:2018}.
\end{abstract}


\section{Introduction}

Double blind randomized controlled trials (RCTs) are seen as the golden standard for causal inference. It is easy to show that the mean difference estimator from well conducted experiments is unbiased (both over random sampling of units into the experiment, and over replicated random treatment allocations among the experimental units). This concept of unbiasedness of an estimator is however often misunderstood as the estimate being `the truth' \citep[cf.][]{Deaton&Cartwright:2018}. In a single experiment the estimate may still be very far from the `true' effect due to an, unfortunate, bad allocation.

With prior information on the assignment process and/or the outcome strategies based on analyzing non-randomized studies may be more efficient than randomization. The conflict between efficiency and bias in the early age of causal inference is well illustrated in \citet{Student:1938}\footnote{The paper was published, with the help of Egon Pearson and Jerzy Neyman, after the death of William Sealy Gosset (i.e. `Student') in 1937.}

\begin{quotation}
I ventured to point out that the advantages of artificial randomization are usually offset by an increased error when compared with balanced arrangements. Prof. Fisher does not agree and has written a paper to test the difference of opinion that there is between us.
\end{quotation}

\noindent The crux with non-randomized analyses is that unbiasedness of an estimator will depend on the prior information and that unbiasedness have priority over efficiency. As researchers and funding bodies have incentives of finding interesting results, the results will be challenged or dismissed (e.g. by journal referees and editors). The most likely reason for the unique position of the double blinded RCT in the research community is that it provides an objective or transparent assessment of the validity of an empirical study, not that it is in any way an efficient strategy of scientific learning.\footnote{For an interesting discussion on decision theory and the motivation for randomization see \citet{Banerjee&etal:2017}.} As false inferences are costly it is thus of importance to use transparent strategies and/or estimators that are more efficient than the mean difference estimator under complete randomized experiments.

As a mean of increasing the efficiency, Fisher suggested blocking or stratification on observed covariates as a method for reducing potential imbalances, and hence bias, in a given randomized trial. An alternative or complement, also suggested by Fisher, is that of rerandomization which was first firmly formalized by \citet{Morgan&Rubin:2012}. The motivation is based on an understanding that, after blocking or stratification, complete randomization within blocks or strata can result in imbalances in other covariates not used in blocking or stratification. In this situation, Fisher is alleged to have recommended rerandomization \citep{Morgan&Rubin:2012}.

The idea with rerandomization is that of removing allocations with imbalance in the observed covariates and then to randomize within the set of allocations with balance in these covariates. With the goal of making inference to the Sample Average Treatment Effect (SATE), \citet{Morgan&Rubin:2012} suggest using the Mahalanobis distance in covariate means of potential treated and control units as the criterion for removing imbalanced allocations. \citet{Morgan&Rubin:2012} show that the mean difference estimator is unbiased and suggest a Fisher randomization test for the inferences to the experiment. The asymptotic distribution of the mean difference estimator was later derived in \citet{Li&etal:2018}, who showed that the asymptotic distribution of the mean difference estimator consist of a linear combination of a normal distributed variable and a truncated normal variable.

This paper discusses several alternative methods of inference under rerandomization.  We first discuss regression adjustment methods, one of which has been shown to have desirable theoretical properties under rerandomization.  We then propose model-based Bayesian inference based on the idea that, with rerandomization, treatment allocation only depends on the covariates of the experimental units and does not depend on the outcomes, so imputation of missing outcomes from the posterior distribution conditional on the covariates provides correct inference.

The next section discusses the Mahalanobis distance based rerandomization procedure of \citet{Morgan&Rubin:2012} and the implication for asymptotic inferences to the SATE. Section 3 discusses regression adjustment methods.  Section 4 proposes model-based Bayesian inference. The small sample performance of various procedures are studied using Monte Carlo simulations in section 5. The paper concludes with a discussion in section 6.

\section{Mahalanobis distance based rerandomization}

\citet{Morgan&Rubin:2012} consider a trial with $n$ units, with $n_{1}$
assigned to treatment and $n_{0}$ assigned to control. Let $W_{i}=1$ or $%
W_{i}=0$ if unit $i$ is assigned treatment or control, respectively, and
define $\bm{W}=(W_{1},...,W_{n})^{\prime }.$ Furthermore let $\bm{X}$
be the $n\times K$ matrix of fixed covariates for the units with the finite population
covariance matrix $\bm{S}_{\bm{x}\bm{x}}$. With $\bm{x}_{i},i=1,...,n,$ the $K\times 1$ covariate vector for unit $i$, this covariance matrix is defined as

\begin{equation}
\bm{S}_{\bm{x}\bm{x}}=\frac{1}{n-1}\sum_{i=1}^n (\bm{x}_i-\bar{\bm{x}})(\bm{x}_i-\bar{\bm{x}})',  \label{eq:Cov}
\end{equation}%

\noindent where $\bar{\bm{x}}=\sum_{i=1}^{n}\bm{x}_{i}/n$.

There are $\tbinom{n}{n_{1}}=n_{A}$ possible treatment allocation (assignment) vectors.  Let $\bm{W}^{j},$ $j=1,...,n_{A}$, denote the $j$th allocation vector, and $\mathbb{W}=(\bm{W}^{1},...,\bm{W}^{n_{A}})$ the complete set of allocations.

The Mahalanobis distance for allocation $j$ is
\begin{equation}
M(\bm{W}^{j}\bm{,X)}= \frac{n}{4}(\widehat{\bm{\tau }}_{X}^{j}{\prime}\bm{S}_{xx}^{-1}\widehat{\bm{\tau }}_{X}^{j}),%
\text{ }j=1,...,n_{A}, \label{eq:MD}
\end{equation}%
\noindent where

\begin{equation*}
\widehat{\bm{\tau }}_{X}^{j}=\frac{1}{n_{1}}\sum_{i=1}^{n}W_{i}^{j}%
\bm{x}_{i}-\frac{1}{n_{0}}\sum_{i=1}^{n}(1-W_{i}^{j})%
\bm{x}_{i}=\overline{\bm{x}}_{1}^{j}-\overline{\bm{x}}%
_{0}^{j},\text{\ }j=1,...,n_{A}.
\end{equation*}%
\citet{Morgan&Rubin:2012} suggested accepting the treatment assignment vector $\bm{W}^{j}$ only when
\begin{equation*}
M(\bm{W}^{j}\bm{,X)}\leq a,
\end{equation*}%
where $a$ is a positive constant. This means that final randomization occur
only within the set
\begin{equation*}
\mathcal{A}_{a}(\bm{X})=\{ \bm{W}^{j}|M(\bm{W}^{j}\bm{,X)}%
\leq a\}.
\end{equation*}
Asymptotically, the Mahalanobis distance follows a $\chi^2_K$ distribution (a chi-square distribution with $K$ degrees of freedom).  Therefore, $a$ can be set to a given quantile of the $\chi^2_K$ distribution.

With well-defined assignment indicator $w=1$ for treatment and $w=0$ for control, define the vector of potential outcome for the $n$ units in the experiment $\bm{Y}(w)=(Y_{1}(w),...,Y_{n}(w))^{\prime }, w=0,1$. For later use, define the individual treatment effect $\tau_i=Y_i(1)-Y_i(0)$, and define the vector $\bm{\tau}_v = (\tau_{1},...,\tau_{n})^{\prime}$. The estimand of interest is the Sample Average Treatment Effect (SATE):
\begin{equation*}
\tau=\frac{1}{n}\sum_{i=1}^n\tau_i=\frac{1}{n}\sum_{i=1}^n Y_i(1)-\frac{1}{n}\sum_{i=1}^n Y_i(0),
\end{equation*}
or $\tau = \bm{1}'\bm{\tau}_v$, where $\bm{1}$ is a column vector of ones.

Let $Y_{i}^{\text{obs}}=Y(W_{i})$ denote the observed outcome for unit $i$, and let
\begin{equation*}
\overline{Y}_{1}^\text{obs}=\frac{1}{n_{1}}\sum_{i=1}^{n}W_iY_{i}^\text{obs} \text{ and }%
\overline{Y}_{0}^\text{obs}=\frac{1}{n_{0}}\sum_{i=1}^{n}(1-W_i)Y_{i}^\text{obs}.
\end{equation*}
Define the standard mean difference estimator for $\tau$ as
\begin{equation}
\widehat{\tau }=\overline{Y}_{1}^\text{obs}-\overline{Y}_{0}^\text{obs}.  \label{eq:MDE}
\end{equation}%

Under the Stable Unit Treatment Value Assumption (SUTVA) \citep{Rubin:1980}, the observed vector of outcomes equals to $\bm{Y}^{\text{obs}} =(Y_{1}(W_1),...,Y_{n}(W_n))^{\prime }$, it is easy to show that the mean difference estimator (\ref{eq:MDE}) is an unbiased estimator for the SATE under complete randomization. Using the fact that the Mahalanobis distance is symmetric in the mean difference of the covariates, \citet{Morgan&Rubin:2012} show that the estimator (\ref{eq:MDE}) also is unbiased under rerandomization when $n_1=n_0=n/2$.

A drawback with the rerandomization strategy is that $\widehat{\tau }$
no longer is normally distributed as is the case under complete randomization. \citet{Li&etal:2018} show that the asymptotic
distribution of $\widehat{\tau }$ after randomly choosing an allocation from
the set $\mathcal{A}_{a}(\bm{X})$ is
\begin{equation}
\lim \sqrt{n}(\widehat{\tau }-\tau )|\widehat{\bm{\tau }}_{X}\overset{d}%
{\rightarrow }\sqrt{V_{\tau }}Q,  \label{eq:Q}
\end{equation}%
\noindent where $V_{\tau }$ is the variance of $\sqrt{n}\widehat{\tau }$ under complete
randomization, $Q=\sqrt{1-R^{2}}\varepsilon _{0}+\sqrt{R^{2}}L_{K,a}$. $\varepsilon _{0}$ is a standard normal variable, which is related to the space orthogonal to that of the covariates and hence is unaffected by rerandomization, $L_{K,a}$ is related to the linear projection of $\bm{Y}(w)$ ($w=0,1$) into the space of covariates and is thus affected by rerandomization, and
\begin{equation*}
R^{2}=\frac{\frac{n}{n_{1}}S_{Y(1)|\bm{x}}^{2}+\frac{n}{n_{0}}S_{Y(0)|%
\bm{x}}^{2}-S_{\tau |\bm{x}}^{2}}{\frac{n}{n_{1}}S_{Y(1)}^{2}+\frac{n}{n_{0}}S_{Y(0)}^{2}-S_{\tau}^{2}}.
\end{equation*}

\noindent Here $S_{Y(w)}$ ($w=0,1$) and $S_{\tau}$ denote, respectively, the finite population variances of $Y_i(w)$ and $\tau_i$,
and $S_{Y(w)|\bm{x}}$ ($w=0,1$) and $S_{\tau|\bm{x}}$ denote, respectively, the finite population variances of the linear projection of $Y_i(w)$
and $\tau_i$ on $\bm{x}_i$. Under homogeneous treatment effects, $S_{\tau |\bm{x}}^{2}=S_{\tau}^{2}=0$. It
follows that $R^{2}=S_{Y(0)|\bm{x}}^{2}/S_{Y(0)}^{2}$, that is, the R-square in a linear regression of $\bm{Y}(0)$ on $\bm{X}$.

The distribution of $L_{K,a}$ has the following form%
\begin{equation*}
L_{K,a}\sim \chi _{K,a}S\sqrt{\beta_K},
\end{equation*}%
where $\chi _{K,a}=\chi _{K}^{2}|\chi _{K}^{2}\leq a$ is a truncated $\chi^2$ random variable, $S$ a random variable
taking values $\pm 1$ with probability 1/2, $\beta_K\sim \text{Beta}(1/2,(K-1)/2)$ is a Beta random
variable degenerating to a point mass at 1 when $K=1$, and ($\chi_{K,a}$,$S$,$\beta_K$) are jointly independent.
\citet{Li&etal:2018} showed that $R^{2}$ can be consistently
estimated using the linear projection of the observed outcomes of the treated and
control units on $\bm{X}$. Together with a consistent estimator for $V_{\tau }$, valid inference can be conducted based on the asymptotic distribution $Q$. In the following, inference conducted under (\ref{eq:Q}) is denoted LDR.

\section{Regression adjustment after rerandomization}

Under complete randomization, \citet{Lin:2013} proposed to estimate $\tau$ using the estimated coefficient on $W_i$ in the OLS regression of $Y_i^{\text{obs}}$ on $W_i$, $\bm{x}_i$ and $W_i(\bm{x}_i-\bar{x})$, and to construct asymptotically valid confidence intervals using the Eicker-Huber-White (EHW) robust standard error estimator \citep{Eicker:1967,Huber:1967,White:1980}.  \citet{Lin:2013} showed that asymptotically, his OLS-adjusted estimator with treatment-covariate interaction is at least as efficient as the mean difference estimator, or the OLS-adjusted estimator without treatment-covariate interaction, i.e., the estimated coefficient on $W_i$ in the OLS regression of $Y_i$ on $W_i$ and $\bm{x}_i$.

Under rerandomization, \citet{Li&Ding:2019} showed that asymptotically \citet{Lin:2013}'s estimator never hurts the precision, and the EHW variance estimator is a convenient approximation to its variance. The theory does not rely on the linear model assumption.

Let $\bm{z}_i$ denote the vector of covariates (including a constant 1) in an OLS regression, the EHW covariance matrix for the regression coefficients is defined as \begin{equation}
\widehat{V}=n\left(\left(\sum_{i=1}^{n}\bm{z}_{i}^{\prime }%
\bm{z}_{i}\right)^{-1}\left(\sum_{i=1}^{n}\widehat{u}_{i}^{2}\bm{z}%
_{i}^{\prime}\bm{z}_{i}\right)\left(\sum_{i=1}^{n}\bm{z}_{i}^{\prime }%
\bm{z}_{i}\right)^{-1}\right), \label{eq:EHW}
\end{equation}
\noindent where $\widehat{u}_{i}$ is the OLS residual.

In an usual OLS regression, the EHW estimator can be severely downward biased in small samples.  A large number of estimators adjusting for this small sample bias has been suggested in the literature (see \citet{MacKinnon:2013} for a review).  Let $\bm{Z}$ denote the matrix of covariates in an OLS regression, and let $\bm{P}=\bm{Z}(\bm{Z}^{\top}\bm{Z})^{-1}\bm{Z}^{\top}$ denote the projection matrix.  The HC2 covariance estimator replaces $\widehat{u}_i$ with $\widehat{\epsilon}_i=\widehat{u}_i/\sqrt{1-h_i}$, where $h_i$ is the $i$th diagonal element of $\bm{P}$.  The HC3 covariance estimator replaces $\widehat{u}_i$ with $\widehat{\epsilon}_i=\widehat{u}_i/(1-h_i)$.  \citet{MacKinnon:2013} pointed out that the HC2 and HC3 estimators correspond to the same asymptotic estimator as the EHW estimator.  The theoretical properties of the HC2 and HC3 estimators under rerandomization are not yet clear.  We will study their performance using a Monte Carlo study in Section~\ref{sec:MonteCarlo}.

\section{Model-based Bayesian inference}

\subsection{The general framework}

Let $Y_{i}^{\text{mis}}=Y(1-W_{i})$ denote the missing potential outcome for unit $i$. We have
\begin{equation*}
\begin{aligned}
\tau&=\frac{1}{n}\left[\sum_{i=1}^n W_iY_i(1)+\sum_{i=1}^n (1-W_i)Y_i(1)\right]-\frac{1}{n}\left[\sum_{i=1}^n W_iY_i(0)+\sum_{i=1}^n (1-W_i)Y_i(0)\right]\\
&=\frac{1}{n}\left[\sum_{i=1}^n W_iY_i^{\text{obs}}+\sum_{i=1}^n (1-W_i)Y_i^{\text{mis}}\right]-\frac{1}{n}\left[\sum_{i=1}^n W_iY_i^{\text{mis}}+\sum_{i=1}^n (1-W_i)Y_i^{\text{obs}}\right].
\end{aligned}
\end{equation*}
\noindent One way of conceptualizing the inference about $\tau$ is that
with $Y_{i}(1)$ and $Y_{i}(0)$ ($i=1,\cdots ,n$) fixed, the distribution of the estimator for $\tau$ stems from the distribution of the prediction error of the missing outcomes. For any allocation vector $\bm{W}$, let $\bm{Y(}\bm{W}\bm{)=(}%
Y_{1}(W_{1}),...,Y_{n}(W_{n}))^{\prime }$. Then the vector of missing outcomes
is $\bm{Y}^{\text{mis}}\bm{=\bm{Y(1-\bm{W}}}\bm{%
\bm{)}}$, and the vector of observed outcomes is $\bm{Y}^\text{obs}\bm{%
=\bm{Y(\bm{W}}}\bm{\bm{)}}$.

Under complete randomization, the probability of treatment allocation vector does not
depend on the potential outcomes or the covariates. We have
\begin{equation*}
\Pr (\bm{W}|\bm{Y}(1),\bm{Y}(0),\bm{X})=\Pr (\bm{W}|\bm{Y}(1),\bm{Y}(0))=\Pr(\bm{W}|\bm{X})=\Pr (%
\bm{W})=1/n_{A}.
\end{equation*}

One way to impute $\bm{Y}^{\text{mis}}$ is to use predictions from the posterior distribution
\begin{eqnarray}
\Pr (\bm{Y}^{\text{mis}}|\bm{Y}^\text{obs},\bm{W})
&=&\frac{\Pr(\bm{Y}^{\text{mis}},\bm{Y}^{\text{obs}},\bm{W})}{\Pr(\bm{Y}^{\text{obs}},\bm{W})}\notag\\
&=&\frac{\Pr(\bm{Y}(1),\bm{Y}(0))\Pr (\bm{W}|\bm{Y}(1),\bm{Y}(0))%
}{\int \Pr(\bm{Y}(1),\bm{Y}(0))\Pr (\bm{W}|\bm{Y}(1),\bm{Y}(0))d\bm{Y}^{\text{mis}}}  \notag \\
&=&\frac{\Pr(\bm{Y}(1),\bm{Y}(0))}{\int \Pr(\bm{Y}(1),\bm{Y}(0))d\bm{Y}^{\text{mis}}}.  \label{eq:Posterior}
\end{eqnarray}%
This implies that we can use a model to characterize the distribution $\Pr(\bm{Y}(1),\bm{Y}(0))$, and impute
$\bm{Y}^{\text{mis}}$ from the model. Suppose that $\bm{\psi}$ is the set of parameters determining $\Pr(\bm{Y}(1),\bm{Y}(0))$. Let $\Pr(\bm{\psi}
|\bm{Y}^\text{obs},\bm{W})$ be the posterior distribution of $\bm{\psi}$. Then the posterior distribution of the
missing outcomes is
\begin{equation*}
\Pr(\bm{Y}^{\text{mis}}|\bm{Y}^{\text{obs}},\bm{W})=\int \Pr(\bm{Y}^{\text{mis}}|\bm{Y}^{\text{obs}},\bm{W},\bm{\psi})\Pr (\bm{\psi}|\bm{Y}^{\text{obs}},\bm{W})d\bm{\psi}.
\end{equation*}

An alternative is to impute $\bm{Y}^{\text{mis}}$ using predictions from the
posterior distribution conditional on $\bm{X}$
\begin{eqnarray}
\Pr(\bm{Y}^{\text{mis}}|\bm{Y}^\text{obs},\bm{W},\bm{X})
&=&\frac{\Pr(\bm{Y}^{\text{mis}},\bm{Y}^{\text{obs}},\bm{W}|\bm{X})}{\Pr(\bm{Y}^{\text{obs}},\bm{W}|\bm{X})}\notag\\
&=&\frac{%
\Pr (\bm{Y(}1\bm{),Y(}0\bm{)}|\bm{X})\Pr (\bm{W}|%
\bm{Y}(1),\bm{Y}(0),\bm{X})}{\int \Pr (\bm{Y(}1\bm{),Y(}0%
\bm{)}|\bm{X})\Pr (\bm{W}|\bm{Y}(1),\bm{Y}(0),%
\bm{X})d\bm{Y}^{\text{mis}}}  \notag \\
&=&\frac{\Pr (\bm{Y(}1\bm{),Y(}0\bm{)}|\bm{X})\Pr (\bm{W}|\bm{X})}{\int \Pr (\bm{Y(}1\bm{),Y(}0\bm{)}|\bm{X}%
)\Pr (\bm{W}|\bm{X})d\bm{Y}^{\text{mis}}}  \notag \\
&=&\frac{\Pr (\bm{Y(}1\bm{),Y(}0\bm{)}|\bm{X})}{\int \Pr (%
\bm{Y(}1\bm{),Y(}0\bm{)}|\bm{X})d\bm{Y}^{\text{mis}}}.\label{eq:Cond_posterior}
\end{eqnarray}
This implies that we can use a model to characterize the distribution $\Pr(\bm{Y}(1),\bm{Y}(0)|\bm{X})$, and impute $\bm{Y}^{\text{mis}}$ from the model. Suppose that $\bm{\theta}$ is the set of parameters determining $\Pr(\bm{Y}(1),\bm{Y}(0)|\bm{X})$. Let $\Pr(\bm{\theta}
|\bm{Y}^{\text{obs}},\bm{W},\bm{X)}$ be the posterior distribution of $\bm{\theta}$. Then the conditional posterior distribution of the
missing outcomes is
\begin{equation*}
\Pr(\bm{Y}^{\text{mis}}|\bm{Y}^\text{obs},\bm{W},\bm{X})=\int \Pr (%
\bm{Y}^{\text{mis}}|\bm{Y}^\text{obs},\bm{W},\bm{X},\bm{\theta }%
)\Pr (\bm{\theta |Y}^\text{obs},\bm{W},\bm{X)}d\bm{\theta }.
\end{equation*}

Under rerandomization in the set $\mathcal{A}_{a}(\bm{X)}$, the
probability of treatment allocation vector depends on the covariates. We have
\begin{equation*}
\Pr(\bm{W}|\bm{Y}(1),\bm{Y}(0),\bm{X})=\Pr (\bm{W}|\bm{X})=1/|\mathcal{A}_{a}(\bm{X)|},
\end{equation*}%
where $|\mathcal{A}_{a}(\bm{X)|}$ denotes the number of elements in $%
\mathcal{A}_{a}(\bm{X)}$. A consequence is that $\Pr (\bm{W}|%
\bm{Y}(1),\bm{Y}(0))$ in the posterior distribution (\ref{eq:Posterior}%
) depends on $\bm{Y}^{\text{mis}}$. This is easily seen from
\begin{eqnarray*}
\Pr(\bm{W}|\bm{Y}(1),\bm{Y}(0)) &=&\int \Pr (\bm{W},%
\bm{X}|\bm{Y}(1),\bm{Y}(0))d\bm{X} \\
&=&\int \Pr (\bm{X}|\bm{Y}(1),\bm{Y}(0))\Pr (\bm{W}|%
\bm{Y}(1),\bm{Y}(0),\bm{X})d\bm{X} \\
&=&\frac{\int \Pr(\bm{X},\bm{Y}(1),\bm{Y}(0))\frac{1}{|\mathcal{A}_{a}(\bm{X})|}d\bm{X}}{\Pr(\bm{Y}(1),\bm{Y}(0))}\\
&=&\frac{\int \Pr (\bm{X})\Pr (\bm{Y}(1),\bm{Y}(0)|\bm{X})%
\frac{1}{|\mathcal{A}_{a}(\bm{X})|}d\bm{X}}{\int \Pr (\bm{X})\Pr
(\bm{Y}(1),\bm{Y}(0)|\bm{X})d\bm{X}},
\end{eqnarray*}%
where $\Pr (\bm{X})$ is the marginal distribution of $\bm{X}$.
This means that \ $\Pr (\bm{W}|\bm{Y}(1),\bm{Y}(0))$ cannot be canceled from the numerator and denominator in the posterior
distribution (\ref{eq:Posterior}) and we can no longer use (\ref{eq:Posterior}) to impute $\bm{Y}^{\text{mis}}$. However, (\ref{eq:Cond_posterior}) still holds, and imputing $\bm{Y}^{\text{mis}}$ based on predictions from the conditional posterior distribution (\ref{eq:Cond_posterior}) is straight forward. Thus, it provides an interesting alternative to \citet{Li&etal:2018}
in making inference about $\tau$ after rerandomization.

Remark: The imputation of $\bm{Y}^{\text{mis}}$ based on predictions from the conditional posterior distribution (\ref{eq:Cond_posterior}) is not contingent on the Mahalanobis distance criterion. This means that model-based inferences for any given rerandomization criterion (see e.g. \citet{Johansson&Schultzberg:2018} for an alternative to the Mahalanobis distance)  can be conducted.

\subsection{Examples of model-based Bayesian inference}

One example of the model for $\Pr (\bm{Y(}1\bm{),Y(}0\bm{)}|\bm{X,\theta})$ is
\begin{equation}
\begin{aligned}
Y_{i}(0)&=\alpha_{0}+\bm{\beta}_{0}^{\prime}\bm{x}_{i}+\varepsilon_{i0},\\
Y_{i}(1)&=\alpha_{0}+\gamma +\bm{\beta}_{0}^{\prime}\bm{x}_{i}+\varepsilon_{i1},
\end{aligned}\label{eq:NointB}
\end{equation}
where $E(\varepsilon_{i0})=E(\varepsilon_{i1})=0$ and $(\varepsilon_{i0},\varepsilon_{i1})$ are independent across units.

We can specify a prior for $\bm{\theta}=(\alpha_{0},\gamma,\bm{\beta}_{0},\sigma_{0}^{2},\sigma_{1}^{2})^{\prime}$, where $\sigma_{w}^{2}=Var(\varepsilon_{iw})$. Following \citet{Imbens&Rubin:2015book} (Section 8.10), we assume that the parameters are independent apriori. The prior distribution for each regression coefficient is normal with mean 0 and variance $100^2$.  The prior distribution for $\sigma_0^2$ or $\sigma_1^2$ is inverse gamma with parameters 1 and 0.01.

Suppose that we have obtained $H$ posterior samples of the parameters.  For each posterior sample, we impute the missing potential outcomes under the unassigned treatment arms, and obtain a posterior sample of $\tau$. In the observed data, there is no information about the correlation coefficient between the two potential outcomes $Y_i(0)$ and $Y_i(1)$ for the same unit. In imputation, we take a conservative approach by assuming that the two potential outcomes are perfectly correlated \citep[][Section 8.6]{Imbens&Rubin:2015book}.

Suppose that the $h$th posterior sample of the parameters is
$$\left(\alpha_0^{(h)},\gamma^{(h)},\bm{\beta}_0^{(h)},\sigma_0^{2(h)},\sigma_1^{2(h)}\right).$$
For units with $W_i=1$, we let $[Y_i(1)]^{(h)}=Y_i^{\text{obs}}$, and impute $Y_i(0)$ as
$$[Y_i(0)]^{(h)}=\alpha_0^{(h)}+[\bm{\beta}_0^{(h)}]'\bm{x}_i+
\frac{\sigma_0^{(h)}}{\sigma_1^{(h)}}(Y_i^{\text{obs}}-\alpha_0^{(h)}-\gamma^{(h)}-[\bm{\beta}_0^{(h)}]'\bm{x}_i).$$
For units with $W_i=0$, we let $[Y_i(0)]^{(h)}=Y_i^{\text{obs}}$, and impute $Y_i(1)$ as
$$[Y_i(1)]^{(h)}=\alpha_0^{(h)}+\gamma^{(h)}+[\bm{\beta}_0^{(h)}]'\bm{x}_i+
\frac{\sigma_1^{(h)}}{\sigma_0^{(h)}}(Y_i^{\text{obs}}-\alpha_0^{(h)}-[\bm{\beta}_0^{(h)}]'\bm{x}_i).$$
The $h$th posterior sample of $\tau$ is
$$\tau^{(h)}=\frac{1}{n}\sum_{i=1}^n\left\{[Y_i(1)]^{(h)}-[Y_i(0)]^{(h)}\right\}.$$

A second example of the model for $\Pr (\bm{Y(}1\bm{),Y(}0\bm{)}|\bm{X,\theta})$ is
\begin{equation}
\begin{aligned}
Y_{i}(0)&=\alpha_{0}+\bm{\beta}_{0}^{\prime}(\bm{x}_{i}-\bar{\bm{x}})+\varepsilon_{i0},\\
Y_{i}(1)&=\alpha_{0}+\gamma +(\bm{\beta}_{0}^{\prime}+\bm{\delta}^{\prime})(\bm{x}_{i}-\bar{\bm{x}})+\varepsilon_{i1},
\end{aligned}\label{eq:IntB}
\end{equation}
where $E(\varepsilon_{i0})=E(\varepsilon_{i1})=0$ and $(\varepsilon_{i0},\varepsilon_{i1})$ are independent across units.

We can specify a prior for $\bm{\theta}=(\alpha_{0},\gamma,\bm{\beta}_{0},\bm{\delta},\sigma_{0}^{2},\sigma_{1}^{2})^{\prime}$, where $\sigma_{w}^{2}=Var(\varepsilon_{iw})$, and impute $\bm{Y}^{\text{mis}}$ from $\Pr(\bm{Y}^{\text{mis}}|\bm{Y}^\text{obs},\bm{W},\bm{X})$.  Details are similar to those for the model in \eqref{eq:NointB}.

\section{Monte Carlo simulation}\label{sec:MonteCarlo}

\subsection{Methods in comparison}

The focus of the Monte Carlo simulation is to compare the performance of the regression adjustment methods and the model-based Bayesian inference to that of the LDR method under Mahalanobis distance based rerandomization.

We use {\it NointE}, {\it NointH2} and {\it NointH3} to denote OLS-adjustment methods without treatment-covariate interaction, with standard errors estimated, respectively, using EHW, HC2 and HC3 estimators.  We use {\it IntE}, {\it IntH2} and {\it IntH3} to denote OLS-adjustment methods with treatment-covariate interaction, with standard errors estimated, respectively, using EHW, HC2 and HC3 estimators.  We use {\it NoinB} and {\it IntB} to denote model-based Bayesian inferences, respectively, under models \eqref{eq:NointB} and \eqref{eq:IntB}.

As a norm we compare the methods to the standard Neyman inference based on the mean difference estimator (denoted `Neyman' in the following text), which is conservative under complete randomization, and has been shown by \citet{Li&etal:2018} to be also conservative under rerandomization.  Specifically, we use the conservative estimator of the standard error of the mean difference estimator in (\ref{eq:MDE}), defined as
$$\sqrt{\frac{2}{n}(s_1^2+s_0^2)},$$
where $s_1^2=\sum_{i=1}^n W_i(Y_i^{\text{obs}}-\bar{Y}_1^{\text{obs}})^2/(n_1-1)$ and $s_0^2 = \sum_{i=1}^n (1-W_i)(Y_i^{\text{obs}}-\bar{Y}_0^{\text{obs}})^2/(n_0-1)$
are the sample variances of the observed outcome for the treatment and control groups.

For Neyman and regression adjustment methods, we consider 95\% confidence intervals in the form of $\hat{\tau}^{[m]}\pm 1.96 \times \widehat{SE}(\hat{\tau}^{[m]})$, where $\hat{\tau}^{[m]}$ is the estimate of $\tau$ based on method $m$, and $\widehat{SE}(\hat{\tau}^{[m]})$ is the estimated standard error of $\hat{\tau}^{[m]}$.  For model-based Bayesian inference, we summarize the posterior samples $\tau^{(h)}$ to obtain 95\% credible intervals for $\tau$. The performance will be studied by comparing the length of a 95\% interval and by comparing the empirical coverage rate to the nominal rate 95\%.

\subsection{Setup of the Monte Carlo simulation}

In each experiment, $n/2$ units are randomly assigned to treatment, and the remaining $n/2$ units are assigned to control. Treatment allocation is randomized until the Mahalanobis distance, given in equation (\ref{eq:MD}), is less than $\chi^2_{K,0.01}$, the 0.01 quantile of the $\chi^2_K$ distribution.

For each unit $i$ ($i=1,\cdots,n$), $\bm{x}_i$ is a $K$ dimensional vector of covariates. The potential outcomes are generated using one of two different data generating processes, DGP1 and DGP2, as follows.

\vspace{0.5cm}

{\bf DGP1:}
\begin{equation*}
\begin{aligned}
Y_i(0)&=\bm{\xi}'\bm{x}_i+\epsilon_{i},\\
Y_i(1)&=Y_i(0)+\lambda+u_{i}.
\end{aligned}
\end{equation*}

\vspace{0.5cm}

{\bf DGP2:}
\begin{equation*}
\begin{aligned}
Y_i(0)&=\bm{\xi}'\bm{x}_i+\epsilon_{i},\\
Y_i(1)&=Y_i(0)+\lambda+\bm{\eta}'(\bm{x}_i-E(\bm{x}))+ u_i.
\end{aligned}
\end{equation*}

\noindent Here, $\bm{\xi}$ is an $K$ dimensional vector with all elements being 1, $\bm{\eta}$ is a $K$ dimensional vector taking values 1, 0.5 and -0.5 if $K=3$ and repeating the values 1, 0.5 and -0.5 in a sequence if $K>3$, $\text{Var}(\epsilon_i)=\sigma_{\epsilon}^2$, where $\sigma_{\epsilon}^2$ is specified such that the super-population squared multiple correlation between $Y(0)$ and $\bm{x}$ equals a given constant $R_0^2$, $\lambda$ is a given constant and $u_i\sim N(0,\sigma_u^2)$, with $\sigma_u^2=c\sigma_{\epsilon}^2$, where $c$ is a given constant.

The following specific settings are considered in the data generation.

\begin{itemize}
\item Sample size: $n\in \{50,100,200,400\}$.
\item The number of covariates: $K\in\{3,10\}$.
\item Each covariate independently follows an $N(0,1)$ or $\exp(1)$ distribution.\footnote{For our setup, $\sigma_{\epsilon}^2=(K(1-R_0^2))/R_0^2$, given that $R_0^2=\text{Var}(\bm{\xi}'\bm{x})/(\text{Var}(\bm{\xi}'\bm{x})+\sigma_{\epsilon}^2)$, and  $\text{Var}(\bm{\xi}'\bm{x})=K$.} When each covariate follows a $N(0,1)$ distribution, we assume that $\epsilon_i \sim \sigma_{\epsilon}N(0,1)$; when each covariate follows an $\exp(1)$ distribution, we assume that $\epsilon_i \sim \sigma_{\epsilon}(\exp(1)-1)$. In both cases, $E(\epsilon_i)=0$ and $Var(\epsilon_i)=\sigma_{\epsilon}^2$.
\item The super-population squared multiple correlation between $Y(0)$ and $\bm{x}$:  $R_0^2 \in \{0.2,0.5\}$.
\item The super-population average treatment effect: $\lambda=0$ or $\lambda=0.3\sqrt{50/n}\sqrt{\text{Var}(Y(0))}$.
\item The ratio of $\sigma_u^2$ to $\sigma_{\epsilon}^2$:
$c\in\{0,0.01,0.1,0.25,0.5\}$. \\Note that under DGP1, $c=0$ corresponds to constant treatment effects across all units.
\end{itemize}

Together with the two data generating processes (DGP's), we are considering a total of seven factors, with a total of 640 different settings. Under each setting, we generate 20 datasets, or samples, with $n$ units.  This means that we in total have 12,800 datasets. For each dataset, we conduct 2000 experiments under rerandomization and calculate the length and the coverage rate of the 95\% intervals, for each of the considered methods.

\subsection{Results}

For each of the two levels on covariate distribution, $K$ and DGP, and the four levels on $n$, we conduct separately an analysis of variance (ANOVA) of six factors: method, $R_0^2$, $\lambda$, $c$, dataset and rerandomized experiment. Details are given in the Appendix.

Table \ref{T:L_ANOVA} displays the percentages of total variation explained by different sources for the length of the 95\% interval, and Table \ref{T:C_ANOVA} displays the corresponding percentages of explained variation for the indicator that the 95\% interval covers the true value of $\tau$.

\begin{table}
\caption{Percentages of different sources of variation for length of interval.}\label{T:L_ANOVA}
\begin{center}
\small
\begin{tabular}{lcccccccc}
\hline\hline
& \multicolumn{2}{c}{$N(0,1)$, $K=3$} & \multicolumn{2}{c}{$exp(1)$, $K=3$}& \multicolumn{2}{c}{$N(0,1)$, $K=10$}& \multicolumn{2}{c}{$exp(1)$, $K=10$}\\
source & DGP1 & DGP2 & DGP1 & DGP2 & DGP1 & DGP2 & DGP1 & DGP2\\
\hline
& \multicolumn{8}{c}{$n=50$}\\
method &  5.01 &  5.01 &  5.32 &  5.32 & 20.63 & 20.63 & 21.93 & 21.93\\
$R_0^2$ & 84.88 & 84.88 & 75.49 & 75.49 & 66.71 & 66.71 & 59.42 & 59.42\\
$\lambda$ &  0.00 &  0.00 &  0.00 &  0.00 &  0.00 &  0.00 &  0.00 &  0.00\\
$c$ &  1.32 &  1.32 &  1.10 &  1.10 &  1.13 &  1.13 &  0.83 &  0.83\\
interaction &  0.68 &  0.68 &  0.52 &  0.52 &  2.77 &  2.77 &  2.84 &  2.84\\
data &  6.68 &  6.68 & 15.68 & 15.68 &  4.67 &  4.67 &  9.77 &  9.77\\
experiment &  1.42 &  1.42 &  1.89 &  1.89 &  4.07 &  4.07 &  5.22 &  5.22\\
\hline
& \multicolumn{8}{c}{$n=100$}\\
method &  3.74 &  3.74 &  3.17 &  3.17 &  6.65 &  6.65 &  6.36 &  6.36\\
$R_0^2$ & 91.30 & 91.30 & 81.93 & 81.93 & 87.42 & 87.42 & 78.49 & 78.49\\
$\lambda$ &  0.00 &  0.00 &  0.00 &  0.00 &  0.00 &  0.00 &  0.00 &  0.00\\
$c$ &  1.23 &  1.23 &  1.31 &  1.31 &  1.24 &  1.24 &  1.33 &  1.33\\
interaction &  0.46 &  0.46 &  0.45 &  0.45 &  0.89 &  0.89 &  0.89 &  0.89\\
data &  2.74 &  2.74 & 12.46 & 12.46 &  2.72 &  2.72 & 11.50 & 11.50\\
experiment &  0.53 &  0.53 &  0.68 &  0.68 &  1.07 &  1.07 &  1.43 &  1.43\\
\hline
& \multicolumn{8}{c}{$n=200$}\\
method &  3.10 &  3.10 &  2.88 &  2.88 &  3.85 &  3.85 &  3.34 &  3.34\\
$R_0^2$ & 92.40 & 92.40 & 87.90 & 87.90 & 91.37 & 91.37 & 87.49 & 87.49\\
$\lambda$ &  0.00 &  0.00 &  0.00 &  0.00 &  0.00 &  0.00 &  0.00 &  0.00\\
$c$ &  1.42 &  1.42 &  1.30 &  1.30 &  1.39 &  1.39 &  1.32 &  1.32\\
interaction &  0.47 &  0.47 &  0.47 &  0.47 &  0.51 &  0.51 &  0.51 &  0.51\\
data &  2.39 &  2.39 &  7.13 &  7.13 &  2.54 &  2.54 &  6.78 &  6.78\\
experiment &  0.22 &  0.22 &  0.31 &  0.31 &  0.34 &  0.34 &  0.57 &  0.57\\
\hline
& \multicolumn{8}{c}{$n=400$}\\
method &  3.36 &  3.36 &  3.23 &  3.23 &  3.29 &  3.29 &  3.21 &  3.21\\
$R_0^2$ & 93.89 & 93.89 & 90.18 & 90.18 & 93.88 & 93.88 & 90.10 & 90.10\\
$\lambda$ &  0.00 &  0.00 &  0.00 &  0.00 &  0.00 &  0.00 &  0.00 &  0.00\\
$c$ &  1.43 &  1.43 &  1.50 &  1.50 &  1.42 &  1.42 &  1.50 &  1.50\\
interaction &  0.45 &  0.45 &  0.45 &  0.45 &  0.44 &  0.44 &  0.49 &  0.49\\
data &  0.76 &  0.76 &  4.49 &  4.49 &  0.84 &  0.84 &  4.47 &  4.47\\
experiment &  0.10 &  0.10 &  0.15 &  0.15 &  0.13 &  0.13 &  0.23 &  0.23\\
\hline
\end{tabular}
\end{center}
\end{table}

\begin{table}
\caption{Percentages of different sources of variation for coverage of interval.}\label{T:C_ANOVA}
\begin{center}
\small
\begin{tabular}{lcccccccc}
\hline\hline
& \multicolumn{2}{c}{$N(0,1)$, $K=3$} & \multicolumn{2}{c}{$exp(1)$, $K=3$}& \multicolumn{2}{c}{$N(0,1)$, $K=10$}& \multicolumn{2}{c}{$exp(1)$, $K=10$}\\
source & DGP1 & DGP2 & DGP1 & DGP2 & DGP1 & DGP2 & DGP1 & DGP2\\
\hline
& \multicolumn{8}{c}{$n=50$}\\
method &  0.50 &  0.50 &  0.62 &  0.62 &  2.77 &  2.77 &  3.15 &  3.15\\
$R_0^2$ &  0.01 &  0.01 &  0.00 &  0.00 &  0.00 &  0.00 &  0.00 &  0.00\\
$\lambda$ &  0.00 &  0.00 &  0.00 &  0.00 &  0.00 &  0.00 &  0.00 &  0.00\\
$c$ &  0.03 &  0.03 &  0.03 &  0.03 &  0.03 &  0.03 &  0.02 &  0.02\\
interaction &  0.03 &  0.03 &  0.04 &  0.04 &  0.02 &  0.02 &  0.03 &  0.03\\
data &  0.08 &  0.08 &  0.35 &  0.35 &  0.08 &  0.08 &  0.27 &  0.27\\
experiment & 99.36 & 99.36 & 98.96 & 98.96 & 97.10 & 97.10 & 96.54 & 96.54\\
\hline
& \multicolumn{8}{c}{$n=100$}\\
method &  0.28 &  0.28 &  0.28 &  0.28 &  0.72 &  0.72 &  0.74 &  0.74\\
$R_0^2$ &  0.00 &  0.00 &  0.00 &  0.00 &  0.00 &  0.00 &  0.00 &  0.00\\
$\lambda$ &  0.00 &  0.00 &  0.00 &  0.00 &  0.00 &  0.00 &  0.00 &  0.00\\
$c$ &  0.03 &  0.03 &  0.03 &  0.03 &  0.03 &  0.03 &  0.03 &  0.03\\
interaction &  0.03 &  0.03 &  0.03 &  0.03 &  0.02 &  0.02 &  0.02 &  0.02\\
data &  0.06 &  0.06 &  0.08 &  0.08 &  0.05 &  0.05 &  0.08 &  0.08\\
experiment & 99.60 & 99.60 & 99.57 & 99.57 & 99.18 & 99.18 & 99.13 & 99.13\\
\hline
& \multicolumn{8}{c}{$n=200$}\\
method &  0.20 &  0.20 &  0.20 &  0.20 &  0.29 &  0.29 &  0.28 &  0.28\\
$R_0^2$ &  0.00 &  0.00 &  0.00 &  0.00 &  0.00 &  0.00 &  0.00 &  0.00\\
$\lambda$ &  0.00 &  0.00 &  0.00 &  0.00 &  0.00 &  0.00 &  0.00 &  0.00\\
$c$ &  0.03 &  0.03 &  0.02 &  0.02 &  0.03 &  0.03 &  0.03 &  0.03\\
interaction &  0.03 &  0.03 &  0.03 &  0.03 &  0.02 &  0.02 &  0.02 &  0.02\\
data &  0.07 &  0.07 &  0.07 &  0.07 &  0.05 &  0.05 &  0.05 &  0.05\\
experiment & 99.66 & 99.66 & 99.68 & 99.68 & 99.61 & 99.61 & 99.62 & 99.62\\
\hline
& \multicolumn{8}{c}{$n=400$}\\
method &  0.22 &  0.22 &  0.21 &  0.21 &  0.18 &  0.18 &  0.17 &  0.17\\
$R_0^2$ &  0.00 &  0.00 &  0.00 &  0.00 &  0.00 &  0.00 &  0.00 &  0.00\\
$\lambda$ &  0.00 &  0.00 &  0.00 &  0.00 &  0.00 &  0.00 &  0.00 &  0.00\\
$c$ &  0.03 &  0.03 &  0.03 &  0.03 &  0.03 &  0.03 &  0.03 &  0.03\\
interaction &  0.03 &  0.03 &  0.02 &  0.02 &  0.02 &  0.02 &  0.02 &  0.02\\
data &  0.05 &  0.05 &  0.05 &  0.05 &  0.04 &  0.04 &  0.04 &  0.04\\
experiment & 99.67 & 99.67 & 99.68 & 99.68 & 99.73 & 99.73 & 99.73 & 99.73\\
\hline
\end{tabular}
\end{center}
\end{table}

From Table \ref{T:L_ANOVA} we can see that most of the variation in the length stems from the different levels of $R_0^2$ (in the range 59.42\% to 93.89\%). The contribution of $R_0^2$ is increasing with $n$, decreasing with $K$ and smaller when the covariates are distributed exponential than normal.\footnote{Examining the main effects of $R_0^2$ (not reported), it is seen that the intervals for $R_0^2=0.5$ are about twice as long as those for $R_0^2=0.2$, irrespective of the number of covariates, the distribution of the covariates, and the data generating process.} The second most important factor is the method. With $n=50$, around 5\% ($K=3$) and 21\% ($K=10$) of the variation is explained by the different methods. When $n=400$, just above 3\% of the variation in the length is explained by the methods, irrespective of the number of covariates, the distribution of the covariates, and the data generating process. Clearly, datasets and experiments have decreasing importance when $n$ increases, and have smaller importance when the covariates are distributed normal than exponential. The datasets explain a maximum of 15.68\% of the variation with exponential distributed covariates and $n=50$, and a minimum of 0.76\% with normal distributed covariates and $n =400$.
The experiments explain a maximum of 5.22\% of the variation with exponential distributed covariates and $n=50$ and a minimum of 0.10\% with normal distributed covariates and $n =400$.

The other factors, including the interactions between method, $R_0^2$, $\lambda$ and $c$, are of minor importance. First, it can be seen that the population treatment effect, $\lambda$, does not contribute to the explanation at all. Second, the heterogenity, $c$, explains 0.83\% to 1.5\% of the variation. The contribution is relatively stable across $n$, covariate distributions, $K$, and DGP. Last, the contribution from the interactions is less than 1\% in all datasets except when $n=50$ and $K=10$, where the interactions explains just below 3\% of the variation in length.

From Table \ref{T:C_ANOVA} we can see that almost all variation in the coverage stems from the experiment (in the range 96.54\% to 99.73\%) and that the contribution increases with the sample size. The second most important factor is the method. For small $n$ the variation explained by the methods increases with $K$. With $n=50$ and $K=10$ around 3\% of the variation in the coverage is explained by the different methods. There is no differences with regard to the covariate distribution for any $n$. The contribution of the datasets, $R_0^2$, $\lambda$, and $c$ and the interactions are zero or of very minor importance.

It is seen that for both the length and the coverage the interactions are of minor importance. This suggest that not much information is lost by summarizing the results on methods from the 12,800 datasets by presenting the main effects from the ANOVA.\footnote{The complete set of the results averaging over the datasets and the experiments can be obtained upon request as 160 tables with four panels by sample size $n$.} That is, we average the length or coverage for each method across $R_0^2$, $\lambda$, $c$, datasets and experiments.

The main effects of methods on length of the interval in percentages of the main effect of the Neyman method is displayed in Table \ref{T:Length}. For $n=400$, the length of the interval for all methods is just above 80\% of the Neyman method for DGP1 and just below 80\% for DGP2. There is no apparent difference in relative lengths across the two covariate distributions and across $K$. With smaller $n$, there are substantial differences between the methods. The overall pattern is that IntE is the most efficient for all $n< 400$, closely followed by LDR and NointE. The least efficient is IntH3, followed by NointH3. For $n =50$ and $K=10$, IntH3 is even less efficient than Neyman.

\begin{table}
\begin{center}
\caption{Main effects of methods on length of interval, measured in percentages of the main effect of Neyman method.}\label{T:Length}
\small
\begin{tabular}{lcccccccc}
\hline\hline
& \multicolumn{2}{c}{$N(0,1)$, $K=3$} & \multicolumn{2}{c}{$exp(1)$, $K=3$}& \multicolumn{2}{c}{$N(0,1)$, $K=10$}& \multicolumn{2}{c}{$exp(1)$, $K=10$}\\
method & DGP1 & DGP2 & DGP1 & DGP2 & DGP1 & DGP2 & DGP1 & DGP2\\
\hline
& \multicolumn{8}{c}{$n=50$}\\
NointE & 79.72 & 76.24 & 78.35 & 73.64 & 72.72 & 68.02 & 73.56 & 68.64\\
IntE & 76.93 & 71.67 & 75.67 & 69.13 & 62.70 & 56.76 & 62.90 & 56.95\\
NointH2 & 84.06 & 80.46 & 82.55 & 77.88 & 83.50 & 78.20 & 84.24 & 78.79\\
IntH2 & 84.03 & 78.28 & 82.67 & 75.50 & 84.05 & 76.12 & 84.91 & 76.86\\
NointH3 & 88.76 & 85.04 & 87.60 & 83.19 & 96.39 & 90.41 & 98.01 & 91.99\\
IntH3 & 92.34 & 86.01 & 93.34 & 85.17 & 116.33 & 105.39 & 125.45 & 113.51\\
NointB & 83.82 & 80.36 & 83.77 & 78.89 & 83.03 & 77.88 & 82.93 & 77.42\\
IntB & 81.95 & 76.35 & 80.28 & 73.34 & 83.84 & 75.91 & 83.21 & 75.34\\
LDR & 78.86 & 73.63 & 77.61 & 71.11 & 68.53 & 63.92 & 68.50 & 63.83\\
\hline
& \multicolumn{8}{c}{$n=100$}\\
NointE & 81.24 & 77.54 & 81.79 & 76.91 & 78.67 & 73.82 & 78.86 & 73.76\\
IntE & 79.94 & 74.31 & 80.42 & 73.42 & 74.10 & 67.33 & 74.25 & 67.25\\
NointH2 & 83.36 & 79.60 & 83.97 & 79.14 & 83.88 & 78.75 & 83.92 & 78.65\\
IntH2 & 83.36 & 77.50 & 83.96 & 76.65 & 83.94 & 76.27 & 84.15 & 76.24\\
NointH3 & 85.57 & 81.75 & 86.35 & 81.63 & 89.54 & 84.12 & 89.62 & 84.19\\
IntH3 & 87.04 & 80.93 & 88.26 & 80.56 & 95.62 & 86.87 & 97.06 & 87.97\\
NointB & 86.89 & 82.64 & 85.81 & 80.84 & 83.75 & 78.90 & 83.51 & 78.14\\
IntB & 81.91 & 76.14 & 83.05 & 75.85 & 82.78 & 75.22 & 82.99 & 75.17\\
LDR & 81.10 & 75.56 & 81.58 & 74.68 & 78.64 & 73.25 & 78.77 & 73.18\\
\hline
& \multicolumn{8}{c}{$n=200$}\\
NointE & 82.46 & 78.55 & 82.59 & 78.11 & 80.86 & 76.23 & 81.91 & 77.05\\
IntE & 81.82 & 75.90 & 81.94 & 75.41 & 78.71 & 72.03 & 79.72 & 72.79\\
NointH2 & 83.51 & 79.58 & 83.66 & 79.21 & 83.40 & 78.64 & 84.44 & 79.50\\
IntH2 & 83.51 & 77.47 & 83.66 & 77.00 & 83.42 & 76.34 & 84.50 & 77.14\\
NointH3 & 84.59 & 80.62 & 84.77 & 80.36 & 86.04 & 81.16 & 87.13 & 82.11\\
IntH3 & 85.27 & 79.10 & 85.56 & 78.74 & 88.52 & 81.00 & 89.91 & 82.08\\
NointB & 85.06 & 81.15 & 86.28 & 81.65 & 82.21 & 78.04 & 84.01 & 78.85\\
IntB & 83.35 & 77.32 & 84.41 & 77.67 & 83.20 & 76.14 & 83.65 & 76.38\\
LDR & 82.58 & 76.77 & 82.69 & 76.29 & 82.81 & 77.48 & 83.58 & 78.02\\
\hline
& \multicolumn{8}{c}{$n=400$}\\
NointE & 82.70 & 78.60 & 83.03 & 78.65 & 82.26 & 77.43 & 82.55 & 77.53\\
IntE & 82.39 & 76.19 & 82.73 & 76.26 & 81.20 & 74.17 & 81.47 & 74.28\\
NointH2 & 83.23 & 79.11 & 83.54 & 79.17 & 83.53 & 78.63 & 83.82 & 78.77\\
IntH2 & 83.23 & 76.96 & 83.55 & 77.01 & 83.54 & 76.31 & 83.83 & 76.44\\
NointH3 & 83.75 & 79.62 & 84.07 & 79.71 & 84.82 & 79.86 & 85.13 & 80.04\\
IntH3 & 84.08 & 77.75 & 84.40 & 77.79 & 85.97 & 78.53 & 86.34 & 78.73\\
NointB & 85.21 & 81.17 & 85.99 & 81.55 & 82.52 & 77.67 & 83.83 & 78.73\\
IntB & 80.55 & 74.49 & 82.51 & 76.06 & 82.59 & 75.44 & 84.69 & 77.22\\
LDR & 82.95 & 76.89 & 83.27 & 76.95 & 85.03 & 79.40 & 85.21 & 79.46\\
\hline
\end{tabular}
\end{center}
\end{table}

The main effect of methods on coverage, measured in percentage differences from the nominal level 95\% is presented in Table \ref{T:CR}. First, note that with $20\times 2000$ replicates (20 datasets and 2000 experiments) the standard error of the mean estimate is 0.11\% ($=100\%\times\sqrt{0.95\times0.05/20/2000}$). This means that a percentage difference of 0.22\% or -0.22\% is statistically significantly different from the nominal level 95\% at the 5\% level.

\begin{table}
\begin{center}
\caption{Main effects of methods on coverage of interval, measured in percentage differences from the nominal level 95\%.}\label{T:CR}
\small
%\footnotesize
\begin{tabular}{lcccccccc}
\hline\hline
& \multicolumn{2}{c}{$N(0,1)$, $K=3$} & \multicolumn{2}{c}{$exp(1)$, $K=3$}& \multicolumn{2}{c}{$N(0,1)$, $K=10$}& \multicolumn{2}{c}{$exp(1)$, $K=10$}\\
method & DGP1 & DGP2 & DGP1 & DGP2 & DGP1 & DGP2 & DGP1 & DGP2\\
\hline
& \multicolumn{8}{c}{$n=50$}\\
Neyman &  2.93 &  3.51 &  3.22 &  3.83 &  2.67 &  3.42 &  2.71 &  3.46\\
NointE & -1.45 & -0.67 & -1.33 & -0.39 & -4.30 & -3.19 & -4.03 & -2.87\\
IntE & -2.43 & -2.44 & -2.40 & -2.33 & -9.96 & -9.96 & -10.64 & -10.58\\
NointH2 & -0.17 &  0.53 & -0.00 &  0.88 & -0.58 &  0.26 & -0.21 &  0.59\\
IntH2 & -0.19 & -0.20 & -0.07 & -0.02 & -0.91 & -0.90 & -0.84 & -0.81\\
NointH3 &  0.96 &  1.56 &  1.28 &  2.08 &  2.12 &  2.65 &  2.52 &  3.01\\
IntH3 &  1.65 &  1.63 &  2.20 &  2.20 &  3.84 &  3.83 &  4.20 &  4.21\\
NointB & -0.22 &  0.53 & -0.44 &  0.45 & -0.89 & -0.01 & -0.64 &  0.19\\
IntB & -0.96 & -0.98 & -1.39 & -1.33 & -0.98 & -0.99 & -1.94 & -1.89\\
LDR & -1.86 & -1.85 & -1.81 & -1.76 & -6.99 & -6.85 & -7.49 & -7.26\\
\hline
& \multicolumn{8}{c}{$n=100$}\\
Neyman &  3.16 &  3.70 &  3.10 &  3.77 &  2.80 &  3.50 &  2.83 &  3.55\\
NointE & -0.42 &  0.27 & -0.49 &  0.35 & -1.46 & -0.49 & -1.36 & -0.39\\
IntE & -0.81 & -0.83 & -0.93 & -0.92 & -3.24 & -3.25 & -3.18 & -3.16\\
NointH2 &  0.15 &  0.81 &  0.13 &  0.95 &  0.12 &  0.94 &  0.15 &  1.01\\
IntH2 &  0.15 &  0.13 &  0.12 &  0.13 &  0.07 &  0.07 &  0.10 &  0.12\\
NointH3 &  0.70 &  1.31 &  0.73 &  1.53 &  1.45 &  2.12 &  1.49 &  2.21\\
IntH3 &  1.03 &  1.02 &  1.14 &  1.15 &  2.45 &  2.45 &  2.63 &  2.65\\
NointB &  1.02 &  1.51 &  0.35 &  1.07 &  0.04 &  0.91 &  0.04 &  0.91\\
IntB & -0.26 & -0.28 & -0.52 & -0.49 & -0.36 & -0.35 & -0.80 & -0.77\\
LDR & -0.56 & -0.57 & -0.67 & -0.65 & -2.37 & -2.35 & -2.37 & -2.31\\
\hline
& \multicolumn{8}{c}{$n=200$}\\
Neyman &  3.22 &  3.74 &  3.27 &  3.85 &  2.80 &  3.48 &  2.68 &  3.43\\
NointE &  0.01 &  0.72 &  0.20 &  0.94 & -0.57 &  0.26 & -0.53 &  0.32\\
IntE & -0.17 & -0.17 &  0.02 &  0.02 & -1.26 & -1.24 & -1.25 & -1.24\\
NointH2 &  0.29 &  0.98 &  0.47 &  1.23 &  0.16 &  0.92 &  0.19 &  0.97\\
IntH2 &  0.29 &  0.28 &  0.47 &  0.49 &  0.14 &  0.13 &  0.17 &  0.19\\
NointH3 &  0.56 &  1.23 &  0.75 &  1.49 &  0.83 &  1.54 &  0.86 &  1.58\\
IntH3 &  0.72 &  0.73 &  0.92 &  0.95 &  1.37 &  1.36 &  1.43 &  1.43\\
NointB &  0.68 &  1.35 &  0.79 &  1.47 & -0.23 &  0.73 &  0.07 &  0.82\\
IntB &  0.06 &  0.05 &  0.28 &  0.30 & -0.02 & -0.00 & -0.52 & -0.51\\
LDR & -0.04 & -0.02 &  0.11 &  0.12 & -1.03 & -1.00 & -1.05 & -1.09\\
\hline
& \multicolumn{8}{c}{$n=400$}\\
Neyman &  3.36 &  3.91 &  3.32 &  3.89 &  2.87 &  3.54 &  2.86 &  3.53\\
NointE &  0.27 &  0.97 &  0.30 &  1.01 &  0.08 &  0.88 &  0.07 &  0.84\\
IntE &  0.19 &  0.18 &  0.22 &  0.22 & -0.23 & -0.19 & -0.24 & -0.23\\
NointH2 &  0.40 &  1.10 &  0.44 &  1.14 &  0.42 &  1.17 &  0.41 &  1.15\\
IntH2 &  0.40 &  0.39 &  0.44 &  0.43 &  0.41 &  0.43 &  0.40 &  0.40\\
NointH3 &  0.52 &  1.22 &  0.57 &  1.28 &  0.73 &  1.46 &  0.73 &  1.46\\
IntH3 &  0.60 &  0.61 &  0.65 &  0.64 &  1.01 &  1.02 &  1.01 &  1.03\\
NointB &  0.87 &  1.57 &  0.99 &  1.62 &  0.08 &  0.86 &  0.63 &  1.35\\
IntB & -0.40 & -0.43 & -0.17 & -0.16 &  0.08 &  0.11 &  0.10 &  0.10\\
LDR &  0.19 &  0.18 &  0.27 &  0.27 & -0.20 & -0.20 & -0.24 & -0.27\\
\hline
\end{tabular}
\end{center}
\end{table}

From Table \ref{T:CR} we can see that for all $n$, the coverage rate of Neyman is around 3\% to 4\% wider than the nominal. The extra coverage is a little bit more substantial for DGP2 than for DGP1. There is no clear pattern across the two covariate distributions nor across $K$. With regard to the coverage of the other methods, it is clear that for $n=50$ there is substantial under-coverage for the most efficient methods displayed in Table \ref{T:Length}. For example, when $K=10$, compared to the nominal level, the coverage is around 10\% to 11\% smaller for IntE, around 7\% smaller for LDR, and around 3\%-4\% smaller for NointE. Also the other methods suffer from some statistically significant under-coverage with the exception of IntH3 and NointH3. For $n=100$ and $n=200$, NointH2, IntH2, NointH3, IntH3 and NointB do not suffer from under-coverage, except that NointB has marginally statistically significant under-coverage with $n=200$, $K=10$, normally distributed covariates and DGP1. For $n=400$ there are still some statistically significant under-coverage for IntE and LDR when $K=10$ and for IntB when $K=3$.

As IntH3 could be no more efficient than Neyman and NointH3 may only marginally improve on the efficiency over Neyman, it seems that there is no method worth pursuing when $n=50$. With small experiments it seems better to make use of the Fisher randomization test which was shown to be efficient in contrast to standard Neyman inference in a small Monte Carlo simulation in \citet{Johansson&etal:2019}.  For $n\geq 100$, IntH2, NointH2 and NointB seem to be useful strategies for inference.

%\bibliographystyle{apalike}
% \bibliography{Simulation}

\section{Discussion}

A given estimate from a randomized experiment can be very far from the `true' mean effect. The reason for the uncertainty is heterogeneity across units in the outcomes. Reducing the heterogeneity by using ex ante information is thus important for the validity of the result from an experiment. Conducting randomization within strata or blocks defined by covariates measure before the experiment is one often used strategy. With many multivalued covariates blocking is not possible, then an alternative or complement is rerandomization. The idea with rerandomization is that of removing allocations with imbalance in the observed covariates and then to randomize within the set of allocations with balance in these covariates. The crux, is that standard asymptotic inference using the mean difference estimator is conservative which if used, invalidate the purpose of the design. Given a Mahalanobis distance criterion for removing imbalanced allocations, suggested by \citet{Morgan&Rubin:2012}, \citet{Li&etal:2018} derived the asymptotic distribution of the mean difference estimator and suggested a consistent estimator of its variance.

This paper focuses on alternative methods for making inference after rerandomization, including regression adjustment methods and model-based Bayesian inference. Results from Monte Carlo simulation are specific to the data generating process chosen and can be hard to generalize. However, based on a large Monte Carlo simulation, we believe it is safe to conclude that several methods work better than the approach in \citet{Li&etal:2018} for small or moderate $n$. For sample sizes as small as $n=50$ we however recommend using the Fisher test. The test has shown to have good performance in contrast to Neyman inference in \citet{Johansson&etal:2019}.

Practitioners may rely on post-experimental regression adjustments, and not consider rerandomization in the first place. This conclusion is not correct however. First, there are efficiency gains for rerandomization due to the common support of the covariates of the treated and controls. Second, we can ensure transparency with the analysis by using the same set of covariates used in rerandomization.

\section*{Appendix}

For each of the two levels on covariate distribution, $K$ and DGP, and the four levels on $n$, we conduct separately an analysis of variance (ANOVA) of six factors: method, $R_0^2$, $\lambda$, $c$, dataset and rerandomized experiment. Details are given below.

For $m$th method, $e$th possible value of $R_0^2$, $f$th possible value of $\lambda$, $g$th possible value of $c$, dataset $d$, and experiment $r$, let $L_{mefg,dr}$ denote the length of the 95\% interval, and let $C_{mefg,dr}$ denote an indicator for whether the 95\% interval covers the true value of $\tau$. In general, let $t_{mefg,dr}$ denote either $L_{mefg,dr}$ or $C_{mefg,dr}$.  Let $M=10$ denote the number of methods, $E=2$ denote the number of possible values of $R_0^2$, $F=2$ denote the number of possible values of $\lambda$, $G=5$ the number of possible values of $c$, $D=20$ denote the number of datasets under each combination of $(m,e,f,g)$, and $R=2000$ denote the number of rerandomized experiments for each dataset.

Let
\begin{equation*}
\begin{aligned}
\bar{t}_{....,..}&=\sum_{m,e,f,g,d,r}t_{mefg,dr}/MEFGDR,\\
\bar{t}_{m...,..}&=\sum_{e,f,g,d,r}t_{mefg,dr}/EFGDR,\\
\bar{t}_{.e..,..}&=\sum_{m,f,g,d,r}t_{mefg,dr}/MFGDR,\\
\bar{t}_{..f.,..}&=\sum_{m,e,g,d,r}t_{mefg,dr}/MEGDR,\\
\bar{t}_{...g,..}&=\sum_{m,e,f,d,r}t_{mefg,dr}/MEFDR,\\
\bar{t}_{mefg,..}&=\sum_{d,r}t_{mefg,dr}/DR,\\
\bar{t}_{mefg,d.}&=\sum_{r}t_{mefg,dr}/R.
\end{aligned}
\end{equation*}
Define the following sums of squares due to different sources of variation:
\begin{equation*}
\begin{aligned}
\text{SS.total}&=\sum_{m,e,f,g,d,r} (t_{mefg,dr}-\bar{t}_{....,..})^2,\\
\text{SS.m}&=EFGDR\sum_{m}(\bar{t}_{m...,..}-\bar{t}_{....,..})^2,\\
\text{SS.e}&=MFGDR\sum_{e}(\bar{t}_{.e..,..}-\bar{t}_{....,..})^2,\\
\text{SS.f}&=MEGDR\sum_{f}(\bar{t}_{..f.,..}-\bar{t}_{....,..})^2,\\
\text{SS.g}&=MEFDR\sum_{g}(\bar{t}_{...g,..}-\bar{t}_{....,..})^2,\\
\text{SS.interaction}&=DR\sum_{m,e,f,g}(\bar{t}_{mefg,..}-\bar{t}_{m...,..}-\bar{t}_{.e..,..}-\bar{t}_{..f.,..}-\bar{t}_{...g,..}+3\bar{t}_{....,..})^2,\\
\text{SS.data}&=R\sum_{m,e,f,g,d}(\bar{t}_{mefg,d.}-\bar{t}_{mefg,..})^2,\\
\text{SS.experiment}&=\sum_{m,e,f,g,d,r}(t_{mefg,dr}-\bar{t}_{mefg,d.})^2,\\
\end{aligned}
\end{equation*}
We have
\begin{equation*}
\text{SS.total}=\text{SS.m}+\text{SS.e}+\text{SS.f}+\text{SS.g}+\text{SS.interaction}+\text{SS.data}+\text{SS.experiment}.
\end{equation*}

In Tables~\ref{T:L_ANOVA} and~\ref{T:C_ANOVA}, we present the percentages of different sources of variation: $100\times\text{SS.m}/\text{SS.total}$, $100\times\text{SS.e}/\text{SS.total}$, $100\times\text{SS.f}/\text{SS.total}$, $100\times\text{SS.g}/\text{SS.total}$, $100\times\text{SS.interaction}/\text{SS.total}$, $100\times\text{SS.data}/\text{SS.total}$, and $100\times\text{SS.experiment}/\text{SS.total}$.  In Table~\ref{T:Length}, we present the main effects of methods on length of interval, measured in percentages of the main effect of Neyman method: $100\times\bar{L}_{m...,..}/\bar{L}_{1...,..}$ for $m=2,\cdots,10$.  In Table~\ref{T:CR}, we present the main effects of methods on coverage of interval, measured in percentage differences from the nominal level 0.95: $100\times(\bar{C}_{m...,..}-0.95)$ for $m=1,\cdots,10$.

\bibliographystyle{apalike}
\bibliography{modelbased}

\begin{thebibliography}{}

\bibitem[Banerjee et~al., 2017]{Banerjee&etal:2017}
Banerjee, A.~V., Chassangx, S., and Snowberg, E. (2017).
\newblock Decision theoretic approaches to experiment design and external
  validity.
\newblock In {\em Handbook of Economic Field Experiments}.
\newblock Chapter 4 Volume 1.

\bibitem[Deaton and Cartwright, 2018]{Deaton&Cartwright:2018}
Deaton, A. and Cartwright, N. (2018).
\newblock {Understanding and misunderstanding randomized controlled trials}.
\newblock {\em Social Science \& Medicine}, 210:2--21.

\bibitem[Eicker, 1967]{Eicker:1967}
Eicker, F. (1967).
\newblock {Limit theorems for regressions with unequal and dependent errors}.
\newblock In {\em Proceedings of the Fifth Berkeley Symposium on Mathematical
  Statistics and Probability}, volume~I, pages 59--82. University California
  Press, Berkeley, CA.

\bibitem[Huber, 1967]{Huber:1967}
Huber, P.~J. (1967).
\newblock {The behavior of maximum likelihood estimates under nonstandard
  conditions}.
\newblock In {\em Proceedings of the Fifth Berkeley Symposium on Mathematical
  Statistics and Probability}, volume~I, pages 221--233. University California
  Press, Berkeley, CA.

\bibitem[Imbens and Rubin, 2015]{Imbens&Rubin:2015book}
Imbens, G.~W. and Rubin, D.~B. (2015).
\newblock {\em Causal inference for statistics, social, and biomedical
  sciences: an introduction}.
\newblock Cambridge University Press.

\bibitem[Johansson and Schultzberg, 2018]{Johansson&Schultzberg:2018}
Johansson, P. and Schultzberg, M. (2018).
\newblock Re-randomization strategies for balancing covariates using
  pre-experimental longitudinal data.
\newblock Working paper 2018:4, Department of Statistics, Uppsala University.

\bibitem[Johansson et~al., 2019]{Johansson&etal:2019}
Johansson, P., Schultzberg, M., and Rubin, D.~B. (2019).
\newblock On optimal re-randomization designs.
\newblock Working paper 2019:3 Department of Statistics, Uppsala University.

\bibitem[Li and Ding, 2019]{Li&Ding:2019}
Li, X. and Ding, P. (2019).
\newblock Rerandomization and regression adjustment.
\newblock To appear in {\it Journal of the Royal Statistical Society, Series
  B}.

\bibitem[Li et~al., 2018]{Li&etal:2018}
Li, X., Ding, P., and Rubin, D.~B. (2018).
\newblock {Asymptotic theory of rerandomization in treatment¨Ccontrol
  experiments}.
\newblock {\em Proceedings of the National Academy of Sciences of the United
  States of America}, 115(37):9157--9162.

\bibitem[Lin, 2013]{Lin:2013}
Lin, W. (2013).
\newblock {Agnostic notes on regression adjustments to experimental data:
  reexamining Freedman's critique}.
\newblock {\em The Annals of Applied Statistics}, 7(1):295--318.

\bibitem[MacKinnon, 2013]{MacKinnon:2013}
MacKinnon, J.~G. (2013).
\newblock {Thirty years of heteroskedasticity-robust inference}.
\newblock In Chen, X. and Swanson, N.~R., editors, {\em Recent Advances and
  Future Directions in Causality, Prediction, and Specification Analysis:
  Essays in Honor of Halbert L. White Jr.}, pages 437--461. Springer, New York.

\bibitem[Morgan and Rubin, 2018]{Morgan&Rubin:2012}
Morgan, K.~L. and Rubin, D.~B. (2018).
\newblock {Rerandomization to improve covariate balance in experiments}.
\newblock {\em Annals of Statistics}, 40(2):1263--1282.

\bibitem[Rubin, 1980]{Rubin:1980}
Rubin, D.~B. (1980).
\newblock {Discussion of ``randomization analysis of experimental data: the
  Fisher random-ization test'' by D. Basu}.
\newblock {\em Journal of the American Statistical Association},
  75(2):591--593.

\bibitem[{Student (Gosset, W. S.)}, 1938]{Student:1938}
{Student (Gosset, W. S.)} (1938).
\newblock {Comparison between balanced and random arrangements of field plots}.
\newblock {\em Biometrika}, 29(3/4):363--378.

\bibitem[White, 1980]{White:1980}
White, H. (1980).
\newblock {Using least squares to approximate unknown regression functions}.
\newblock {\em International Economic Revi}, 21(1):149--170.

\end{thebibliography}

\end{document}